# Chapter 1
# Microservices Anti-Patterns: A Taxonomy

Davide Taibi, Valentina Lenarduzzi, Claus Pahl

**Abstract** Several companies are re-architecting their monolithic information sys-
tems with microservices. However, many companies migrated without experience
on microservices, mainly learning how to migrate from books or from practitioners'
blogs. Because of the novelty of the topic, practitioners and consultancy are learning
by doing how to migrate, thus facing several issues but also several benefits. In this
chapter, we introduce a catalog and a taxonomy of the most common microservices
anti-patterns in order to identify common problems. Our anti-pattern catalogue is
based on the experience summarized by different practitioners we interviewed in
the last three years. We identified a taxonomy of 20 anti-patterns, including orga-
nizational (team oriented and technology/tool oriented) anti-patterns and technical
(internal and communication) anti-patterns. The results can be useful to practitioners
to avoid experiencing the same difficult situations in the systems they develop. More-
over, researchers can benefit of this catalog and further validate the harmfulness of
the anti-patterns identified.

## 1.1 Introduction

Microservices are increasing in popularity, being adopted by several companies
including SMEs but also big players such as Amazon, LinkedIn, Netflix, and Spotify.
   Microservices are small and autonomous services deployed independently, with
a single and clearly defined purpose [1][4]. Microservices propose to vertically

Davide Taibi
Tampere University e-mail: davide.taibi@tuni.fi

Valentina Lenarduzzi
Tampere University e-mail: valentina.lenarduzzi@tuni.fi

Claus Pahl
Free University of Bozen-Bolzano e-mail: claus.pahl@unibz.it

decompose the applications into a subset of business-driven independent services. Every service can be developed, deployed and tested independently by different development teams, and by means of different technology stacks. Microservices have a lot of advantages. They can be developed in different programming languages, they can scale independently from other services, and they can be deployed on the hardware that best suits their needs. Moreover, because of their size, they are easier to maintain and more fault-tolerant since the failure of one service may not break the whole system, which could happen in a monolithic system.

However, the migration to microservice is not an easy task [5][6][20][29]. Companies commonly start the migration without experience with microservices, only in few cases hiring a consultant to support them during the migration [6][20]. Therefore, companies often face common problems, which are mainly due to their lack of knowledge regarding bad practices and patterns [5][6][20][21].

In this work, we provide a taxonomy of architectural and organizational anti-patterns specific to microservices-based systems, together with possible solutions to overcome them. To produce this catalog, we adopted a mixed research method, combining industrial survey, literature review and interviews. We replicated and extended our previous industrial surveys [6][7] also considering the bad practices proposed by practitioners (Table 1.7). We surveyed and interviewed 27 experienced developers in 2018, focusing on bad practices they found during the development of microservices-based systems and the solutions they adopted to overcame them. The interviews were based on the same questionnaire we adopted in [6], with the addition of a section where we asked the interviewees if they experienced some of the anti-patterns proposed by practitioners (Table 1.7). We proposed a taxonomy of 20 microservices-specific anti-patterns, by applying an open and selective coding [3] procedure to derive the anti-pattern catalog from the practitioners' answers.

The goal of this work is to help practitioners avoid these bad practices altogether or deal with them more efficiently when developing or migrating monoliths to microservices-based systems.

The remainder of this chapter is structured as follows. Section 1.2 describe the empirical study we carried out. Section 1.3 reports results. Section 1.4 describe the background on microservices and related works. While Section 1.5 draws conclusions.

## 1.2  The Empirical Study

As reported in the introduction, the goal of this work is to provide a taxonomy of anti-patterns specific for microservices.

We first collected the anti-patterns by means of a survey among experienced developers, collecting bad practices in microservices architectures and how they overcame them. Then, we classified the anti-patterns and proposed a taxonomy. Therefore, we formulated our research questions as:

**RQ1  What anti-patterns have been experienced by practitioners when using microservices?**
In this RQ, we aim at understanding if practitioners experienced some anti-patterns, including these proposed in previous works (Table 1.7), which problem the anti-pattern caused and how they overcome to the problem they caused.

**RQ2  What type of anti-patterns have been identified by practitioners?**
In this RQ, we aim at classify the different anti-patterns identified by means of a taxonomy.

### 1.2.1  Study Design

We designed the survey with semi-structed interviews, both in a structured fashion, via a questionnaire with closed answers, and in a less structured way, by asking the interviewees open-answer questions to elicit additional relevant information (such as possible issues when migrating to microservices). One of the most important goals of the questionnaire was to understand which bad practices have the greatest impact on system development and which solutions are being applied by practitioners to overcome them.

Thus, we asked the interviewees to rank every bad practice on a scale from 0 to 10, where 0 meant "the bad practice is not harmful" and 10 meant "the bad practice is exceedingly harmful". Moreover, we clarified that only the ranking of the bad practices has real meaning. As an example, a value of 7 for the "Hardcoded IPs" bad practice and 5 for "Shared Persistence" shows that Hardcoded IPs is believed to be more harmful than Shared Persistence, but the individual values of 7 and 5 have no meaning in themselves. Harmful practice refers to a practice that has created some issue for the practitioner, such as increasing maintenance effort, reducing code understandability, increasing faultiness, or some other issue.

The interviews were based on a questionnaire organized into four sections, according to the information we aimed to collect:

- Personal and company information: interviewee's role and company's application domain.
- Experience in developing microservices-based systems: number of years of experience in developing microservices. This question was asked to ensure that data was collected only from experienced developers.
- Microservices bad practices harmfulness: List of the practices that created some issues during the development and maintenance of microservices-based applications, ranked according to their harmfulness on a 10-point Likert scale. Moreover, for each practice, we asked to report which problem generated and why they considered as harmful. For this answer, the interviewer did not provide any hints, letting the participants report the bad practices they had faced while developing or maintaining microservices-based systems. Moreover, in order to avoid influencing the interviewees, we asked them to list their own bad practices, without providing them with a list of pitfalls previously identified by practitioners [6,9,10,11,12].
- Bad practices solutions: For each bad practice identified, we asked the participants to report how they overcame it.

- Rank the harmfulness of the bad practices previously identified in previous study [6] and those identified by practitioners (Table 1.7): after the open questions, for each of the bad practices reported we asked 1) if they ever experienced that issue, 2) in case they used, to rank the harmfulness of them on a 10-point Likert scale. We decided to ask for the ranking of the harmfulness of the bad practices proposed in the literature after the open questions, to avoid to bias the interviewees with the results of the previous questionnaire. While ranking the bad practices proposed in the literature, practitioners also noted if some of the bad practices they specified in the open questions had the same meaning as those reported in the literature, thus reducing the risk of misinterpretation of their classification.

We are aware that the structure of this questionnaire increased the collection time, but helped us to increase the quality of the answers avoiding to bias the participants with a preselected set of answers.

### 1.2.2 Study Execution

All interviews were conducted in person. An online questionnaire might have yielded a larger set of answers, but we believe that face-to-face interviews are more reliable for collecting unstructured information, as they allow establishing a more effective communication channel with the interviewees and make it easier to interpret the open answers.

The interviewees were asked to provide individual answers, even if they worked in the same group. This allowed us to get a better understanding of different points of view, and not only of the company point of view.

We selected the participants from the attendees of two practitioner events. We interviewed 14 participants of the O'Reilly Software Architecture Conference in London (October 2018) and 13 participants of the international DevOps conference in Helsinki (December 2018). During the interviews, we first introduced our goals to the participants and then asked them if they had at least two years of experience in developing microservices-based systems, so as to save time and avoid including non-experienced practitioners.

### 1.2.3 Data analysis

We partitioned the responses into homogeneous subgroups based on demographic information in order to compare the responses obtained from all the participants with the different subgroups separately.

Ordinal data, such as 10-point Likert scales, were not converted into numerical equivalents, since using a conversion from ordinal to numerical data entails the risk that subsequent analysis will give misleading results if the equidistance between the values cannot be guaranteed. Moreover, analyzing each value of the scale allows us to better identify the possible distribution of the answers. The harmfulness of the bad practices was analyzed calculating the medians, as customary done for ordinal ranges.

Open questions were analyzed via open and selective coding [3]. In addition, practitioners were asked to report if some of the bad practice they reported in the first section of the questionnaire were related to some of the anti-patterns reported in Table 1.7, some practitioners proposed a set of bad practices not directly related to the existing anti-patterns. Therefore, for these cases, we extracted codes from the answers provided by the participants and answers were grouped into different anti-patterns. Answers were interpreted extracting concrete sets of similar answers and grouping them based on their perceived similarity. The qualitative data analysis has been conducted individually by two authors. Moreover, in order to get a fair/good agreement on the first iteration of this process, pairwise inter-rater reliability was measured across the three sets of decisions. Based on the disagreements, we clarified possible discrepancies and different classifications. A second iteration resulted in 100% agreement among all the authors.

The taxonomy was then proposed by two of the authors that grouped different set of anti-patterns into homogeneous categories and then was validated by the third author.

## 1.3 The Study results

We conducted 27 interviews with participants belonging to 27 different organizations. No unexperienced participants such as students, academics, or non-industrial practitioners were considered for the interviews. 36 % of our participants were software architects, 19% project managers, 38% experienced developers, 7% agile coaches. All the interviewees had at least four years of experience in software development. 28.57% of our interviewees worked for software development companies, 28.57% for companies that produce and sell only their own software as a service (e.g., website builders, mobile app generators, and others), and 9.52% in banks/insurances. 17% had adopted microservices for more than five years, 60% had adopted them for three to four years, and the remaining 23% for two to three years.

On top of the proposed bad practices identified in [6] and in (Table 1.7), the practitioners reported a total of 9 different bad practices together with the solutions they had applied to overcome them. Two authors of this paper grouped similar practices (considering both the description and the justification of the harmfulness provided by the participants) by means of open and selective coding [3]. In cases where they interpreted the descriptions differently, they discussed incongruities so as to achieve agreement on similar issues.

### 1.3.1 Data Analysis and Interpretation

The answers were mainly analyzed using descriptive statistics. No noticeable differences emerged among different roles or domains. As reported in Table 1.1, eight anti-patterns proposed by practitioners have never been experienced by the interviewees while four new ones were introduced. Wrong Cuts, Cyclic Dependencies, Hardcoded Endpoints, and Shared Persistency are still considered the most harmful issues.

Differently from our previous study, more organizational issues are now playing an important role during the migration to microservices. Participants considered very important the alignment between the organization structure and the system architecture. Moreover, they also highlighted the importance of having a fully functional DevOps tools pipeline, including continuous testing, integration and delivery.

However, not all the anti-patterns proposed by practitioners resulted as being harmful. As an example, the shared ownership of several microservices from one time is not considered as very important.

Table 1.1 lists the microservices anti-patterns together with the number and percentage of practitioners who mentioned them (column Answer # and %) and the median of the perceived harmfulness reported.

We identify the taxonomy classifying the anti-patterns experienced by our interviewees into two groups: technical and organizational anti-patterns. Figure 1 depicts the proposed classification. For matter of completeness, we report (underlined) the anti-patterns proposed by the practitioners (Table 1.7) but never experienced by our interviewees. In Table 1.3, Table 1.2 and Table 1.4 we describe the technical anti-patterns that have been reported by our interviewees, and the solutions they adopted to overcome to the issues they generated. In Table 1.5 and Table 1.6 we describe the organizational anti-patterns identified. The results of this work are subject to some threats to validity, mainly due to the selection of the survey participants and to the data interpretation phase. Different respondents might have provided a different set of answers. To mitigate this threat, we selected a relatively large set of participants working in different companies and different domains. During the survey, we did not propose a predefined set of bad practices to the participants; therefore, their answers are not biased by the results of previous works. However, as the surveys were carried out during public events, we are aware that some participants may have shared some opinions with others during breaks and therefore some answers might have been partially influenced by previous discussions. Finally, the answers were aggregated independently by the two authors by means of open and selective coding [3].

- Technical
    - Internal: Anti-patterns that impact the individual microservice.
        - API Versioning
        - Hardcoded Endpoints
        - Inappropriate Service Intimacy
        - Megaservice
        - Local Logging
        - Lack of service abstraction
    - Communication: Anti-patterns related to the communication between microservices
        - Cyclic Dependency
        - ESB Usage
        - No API-Gateway
        - Shared Libraries
        - Timeout
    - Others
        - Lack of Monitoring
        - Shared Persistence
        - Wrong Cuts
- Organizational
    - Team-Oriented: Anti-patterns related to the team's dynamics.
        - Legacy Organization
        - Non-homogeneous adoption
        - Common Ownership
        - Microservice Greedy
        - Magic Pixie Dust
        - Microservice as the goal
        - Pride
        - Sloth
    - Technology and Tool Oriented
        - Focus on latest technologies
        - Lack of Microservice Skeleton
        - No DevOps tools

**Fig. 1.1** The Proposed Microservice Anti-Pattern Taxonomy. The anti-patterns underlined were proposed by the practitioners (Table 1.7) but never experienced by our interviewees.

**Table 1.1** The microservices anti-patterns identified in the survey

| Microservices Anti-Pattern | Also proposed by | Answers # | Answers % | Perceived Harmfulness (0-10) |
|---|---|---|---|---|
| Hardcoded Endpoints | [6][10] | 10 | 37 | 8 |
| Wrong Cuts | [6] | 15 | 56 | 8 |
| Cyclic Dependency | [6] | 5 | 19 | 7 |
| API Versioning | [8][10] | 6 | 22 | 6.05 |
| Shared Persistence | [6][9][11] | 10 | 37 | 6.05 |
| ESB Usage | [6] | 2 | 7 | 6 |
| Legacy Organization | [22] | 2 | 7 | 6 |
| Local Logging | NEW | 17 | 63 | 6 |
| Megaservice | [9] | 5 | 19 | 6 |
| Inappropriate Service Intimacy | [6] | 5 | 19 | 5 |
| Lack of Monitoring | NEW | 3 | 11 | 5 |
| No API-Gateway | [6][12] | 4 | 15 | 5 |
| Shared Libraries | [6][8] | 8 | 30 | 4 |
| Too Many Technologies | [6][13][22] | 3 | 11 | 4 |
| Lack of Microservice Skeleton | NEW | 9 | 33 | 3.05 |
| Microservice Greedy | [6][13][22] | 4 | 15 | 3 |
| Focus on Latest Technologies | [22] | 2 | 7 | 2.05 |
| Common Ownership | [13] | 4 | 15 | 2 |
| No DevOps Tools | NEW | 2 | 7 | 2 |
| Non-homogeneous adoption | [22] | 2 | 7 | 2 |
| Lack of service abstraction | [9] | 0 | | |
| Magic Pixie dust | [22] | 0 | | |
| Microservices as the goal | [22] | 0 | | |
| Pride | [13] | 0 | | |
| Sloth | [13] | 0 | | |
| Timeout | [8][10] | 0 | | |
| Try to fly before you can walk | [22] | 0 | | |

Harmfulness was measured on a 10-point Likert scale,
0 means "the bad-practice is not harmful" and 10 means "the bad-practice is extremely harmful".

**Table 1.2** Internal Anti-patterns

| Microservices Anti-Pattern | Description (Desc) / Detection (Det) | Problem it may cause (P) / Adopted Solutions (S) |
|---|---|---|
| API Versioning | **Desc:** APIs are not semantically versioned.<br><br>**Det:** Lack of semantic versions in APIs (e.g., v1.1, 1.2 )<br><br>Also proposed as "Static Contract Pitfall" by Richards [8] and Saleh [10]. | **P:** In the case of new versions of non-semantically versioned APIs, API consumers may face connection issues. For example, the returning data might be different or might need to be called differently.<br>**S:** APIs need to be semantically versioned to allow services to know whether they are communicating with the right version of the service or whether they need to adapt their communication to a new contract. |
| Hardcoded Endpoints | **Desc/Det:** Hardcoded IP addresses and ports of the services between connected microservices.<br>Also proposed by Saleh [10] as "Hardcoded IPs and Ports". | **P:** Microservices connected with hardcoded endpoints lead to problems when their locations need to be changed.<br><br>**S:** Adoption of a service discovery approach. |
| Inappropriate Service Intimacy | **Desc:** The microservice keeps on connecting to private data from other services instead of dealing with its own data.<br>**Det:** Request of private data of other microservices. Direct connection to other microservices databases. | **P:** Connecting to private data of other microservices increases coupling between microservices. The problem could be related to a mistake made while modeling the data.<br>**S:** Consider merging the microservices. |
| Megaservice | **Desc:** A service that does a lot of things. A monolith.<br>**Det:** Several business processes implemented in the same service. Service composed by several modules, and developed by several developers, or several teams | **P:** The same problem of a monolithic system<br><br>**S:** Decompose the megaservice into smaller microservices |
| Local Logging | **Desc/Det:** Logs are stored locally in each microservice, instead of using a distributed logging system | **P:** Errors and microservices information are hidden inside each microservice container. The adoption of a distributed logging system eases the monitoring of the overall system |

**Table 1.3** Communications Anti-patterns

| Microservices Anti-Pattern | Description (Desc) / Detection (Det) | Problem it may cause (P) / Adopted Solutions (S) |
|---|---|---|
| Cyclic Dependency | **Desc:** A cyclic chain of calls between microservices | **P:** Microservices involved in a cyclic dependency can be hard to maintain or reuse in isolation. |
|  | **Det:** Existence of cycles of calls between microservices. E.g., A calls B, B calls C, and C calls back A. | **S:** Refinement of the cycles according to their shape [15] and application of an API-Gateway pattern [4]. |
| ESB Usage | **Desc/Det:** The microservices communicate via an Enterprise Service Bus (ESB). | **P:** ESB adds complexities for registering and deregistering services on the ESB. |
|  | Usage of ESB for connecting microservices | **S:** Adopt a lightweight message bus instead of the ESB. |
| No API-Gateway | **Desc:** Microservices communicate directly with each other. In the worst case, the service consumers also communicate directly with each microservice, increasing the complexity of the system and decreasing its ease of maintenance. | **P:** Our interviewees reported being able to work with systems consisting of 50 interconnected microservices; however, if the number was higher, they started facing communication and maintenance issues. |
|  | **Det:** Direct communication between microservices | **S:** Application of an API-Gateway pattern [4] to reduce the communication complexity between microservices. |
|  | Also proposed by Alagarasan [12] as "Not having an API-Gateway". |  |
| Shared Libraries | **Desc/Det:** Usage of shared libraries between different microservices. | **P:** Tightly couples microservices together, leading to a loss of independence between them. Moreover, teams need to coordinate with each other when they need to modify the shared library. |
|  | Also named "I was taught to share" by Richards [8] | **S:** Two possible solutions: 1) accept the redundancy to increase dependency among teams; 2) extract the library to a new shared service that can be deployed and developed independently by the connected microservices. |

**Table 1.4** Other Technical Anti-patterns

| Microservices Anti-Pattern | Description (Desc) / Detection (Det) | Problem it may cause (P) / Adopted Solutions (S) |
|---|---|---|
| Lack of Monitoring | **Desc/Det:** Lack of usage of monitoring systems, including systems to monitor if a service is alive or if it responds correctly | **P:** A service could be offline, and developers could not realize it without continuously check manually<br><br>**S:** Adoption of a monitoring system |
| Shared Persistence | **Desc/Det:** Different microservices access the same relational database. In the worst case, different services access the same entities of the same relational database.<br>Also proposed by Bogard as "data ownership" [11]. | **P:** This anti-pattern highly couples the microservices connected to the same data, reducing team and service independence.<br><br>**S:** Three possible solutions for this anti-pattern are: use 1) independent databases for each service, 2) a shared database with a set of private tables for each service that can be accessed only by that service, 3) a private database schema for each service. |
| Wrong Cuts | **Desc:** Microservices should be split based on business capabilities, not on technical layers (presentation, business, data layers). | **P:** Wrong separation of concerns, increased data-splitting complexity.<br><br>**S:** Clear analysis of business processes and the need for resources. |

**Table 1.5** Organizational (Team-Oriented) Anti-patterns

| Microservices Anti-Pattern | Description (Desc) / Detection (Det) | Problem it may cause (P) / Adopted Solutions (S) |
|---|---|---|
| Legacy Organization | **Desc:** The company still work without changing their processes and policies. As example, with independent Dev and Ops teams, manual testing and scheduling common releases.<br>Also proposed as "Red Flag" by Richardson [22]. | **P:** Developers are bound to the traditional process, they cannot benefit of most of the benefits of microservices. |
| Non-homogeneous adoption | **Desc/Det:** Only few teams migrated to microservices, and the decision if migrate or not is delegated to the teams.<br>Also defined as âĂIJscattershot adoptionâĂİ by Richardson [22]. | **P:** Duplication of effort. E.g. effort for building the infrastructure, deployment pipelinesâĂę |
| Common Ownership | **Desc/Det:** One team own all the microservices. | **P:** Each microservice will be developed in pipeline, and the company is not benefiting of the development independency. |
| Microservice Greedy | **Desc:** Teams tend to create of new microservices for each feature, even when they are not needed. Common examples are microservices created to serve only one or two static HTML pages.<br>**Det:** Microservices with very limited functionalities (e.g., a microservice serving only one static HTML page) | **P:** This anti-pattern can generate an explosion of the number of microservices composing a system, resulting in a useless huge system that will easily become unmaintainable because of its size. Companies should carefully consider whether the new microservice is needed. |

**Table 1.6** Organizational (Technology and Tool Oriented) Anti-patterns

| Microservices Anti-Pattern | Description (Desc) / Detection (Det) | Problem it may cause (P) / Adopted Solutions (S) |
|---|---|---|
| Focus on latest technologies | **Desc:** The migration is focused on the adoption of the newest and coolest technologies, instead of based on real. The decomposition is based on the needs of the different technologies aimed to adopt. Also proposed as "Focusing on Technology" by Richardson [22]. | **P:** The development is not solving existing problems but is mainly following the technology vendor recommendations. |
| Lack of Microservice Skeleton | **Desc/Det:** Each team develop microservices from scratch, without benefit of a shared skeleton that would speed-up the connection to the shared infrastructure (e.g. connection to the API-Gateway) | **P:** Developers have to re-develop the skeleton from scratch every time, wasting time and increasing the risk of errors |
|  |  | **S:** introduction of a common code boilerplate |
| No DevOps tools | **Desc:** The company does not employ CD/CI tools and developers need to manually test and deploy the system | P: Slower productivity, possible deployment errors due to the lack of automation. |
| Too Many Technologies | **Desc/Det:** Usage of different technologies, including development languages, protocols, frameworks... Also proposed by Bryant [13] as "Lust" and "Gluttony". | **P:** The company does not define a common policy. Although microservices allow the use of different technologies, adopting too many different technologies can be a problem in companies, especially in the event of developer turnover. Companies should carefully consider the adoption of different standards for different microservices, without following new hypes. |

## 1.4 Background and Related Works

Microservices are relatively small and autonomous services deployed independently, with a single and clearly defined purpose [1]. Because of their independent deployment, they have a lot of advantages. They can be developed in different programming languages, they can scale independently from other services, and they can be deployed on the hardware that best suits their needs. Moreover, because of their size, they are easier to maintain and more fault-tolerant since a failure of one service will not break the whole system, which could happen in a monolithic system. Since every microservice has its own context and set of code, each microservice can change its entire logic from the inside, but from the outside, it still does the same thing, reducing the need of interaction between teams[32][33].

Different microservice patterns have been proposed by practitioners [28] and researchers[16]. Zimmerman et al [28] proposed a joint collaboration between academia and industry to collect microservices patterns. However, all these works focus on patterns that companies should follow when implementing microservices-based systems instead of anti-patterns and bad smells to avoid. Balalaie [21] also conducted an industrial survey to understand how companies migrated to microservices, obtaining 15 migration patterns.

As for anti-patterns, several generic architectural anti-pattern have been defined in the last years in different research works [2, 17, 18, 19] and different tools have been proposed both from industry and from researchers to detect them [24, 25, 26, 30]. However, to the best of our knowledge, no peer-reviewed work and, in particular, only few empirical studies have proposed bad practices, anti-patterns, or smells specifically concerning microservices. On the other side, practitioners proposed several anti-patterns, mainly by means of talks in technical events.

As for research works, Bogner et al. [31] reviewed microservices bad smells and anti-patterns proposed in the literature, extracting 36 anti-patterns from 14 peer-reviewed publications. Their survey includes the vast majority of anti-patterns and bad smells highlighted also by practitioners. However, they did not report or classify their harmfulness.

In our previous study [6], we performed an industrial survey investigating the migration processes adopted by companies to migrate to microservice. One of the side results, was that practitioners are not aware about the patterns they should adopt and about the anti-patterns to avoid. In another work we investigated the most used architectural patterns [16] while finally in our latest work [7][14] we investigated "bad smells" of microservices, specific to systems developed using a microservice architectural style, together with possible solutions to overcome these smells. We identified 20 microservice-specific organizational and technical anti-patterns, bad practices that practitioners found during the development of microservice-based systems, and we highlighted how practitioners overcome to that bad practices interviewing 72 experienced developers. Our results [7] are also confirmed by a recent industrial survey performed by Soldani et al. [20]. They identified, and taxonomically classified and comparing the existing grey literature on pains and gains of microservices, from their design to their development, among 51 industrial studies. Based on the results, they prepared a catalog of migration and rearchitecting patterns, in order to facilitate rearchitecting non cloud-native architectures during the migration to a cloud-native microservices-based architecture. In another study [23] we proposed a decomposition framework to decompose monolithic systems into microservices, where one of the most important steps, is the investigation and the removal of possible microservices anti-patterns.

Balalaie [21] also performed an industrial survey proposing a set of 15 migration patterns to understand how companies migrated to microservices. However, they did not report bad practices or anti-patterns. Practitioners have started to discuss bad practices in microservices in recent years. In his eBook [8], Richards introduced three main pitfalls: "Timeout", "I Was Taught to Share", and "Static Contract Pitfall". Moreover, in the last two years, practitioners have given technical talks about bad practices they experienced when building microservices. In Table 1.7, we summarize the main bad practices presented in these works. Chris Richardson recently gave a talk on microservices anti-patterns [22] proposing six organizational anti-patterns based on his consultancy experience.

**Table 1.7** The main pitfalls proposed in non-peer reviewed literature and practitioner talks

| Bad Practice | Description |
|---|---|
| **Timeout** (Richards [8]) **Dogpiles** (Saleh [10]) | Management of remote process availability and responsiveness. Richards recommends using a timeout value for service responsiveness or sharing the availability and the unavailability of each service through a message bus, so as to avoid useless calls and potential timeout due to service unresponsiveness. |
| **I Was Taught to Share** (Richards [8]) | Sharing modules and custom libraries between microservices |
| **Static Contract Pitfall** (Richards [8], Saleh [10]) | Microservices API are not versioned and therefore service consumers may connect to older versions of the services. |
| **Mega-Service** (Shoup [9]) | A service that is responsible for many functionalities and should be decomposed into separated microservices |
| **Shared Persistence** (Shoup [9]) **Data Ownership** (Bogard [11]) | Usage of shared data among services that access the same database. Microservices should own only the data they need and possibly share them via APIs. |
| **Leak of Service Abstraction** (Shoup [9]) | Service interfaces designed for generic purposes and not specifically designed for each service |
| **Hardcoded IPs and Ports** (Saleh [10]) | Hardcoding the IP address and ports of communicating services, therefore making it harder to change the service location afterwards |
| **Not having an API-Gateway** (Alagarasan [12]) | Exposing services through an API-Gateway layer and not connecting them directly so as to simplify the connection, supporting monitoring, and delegating authorization issues to the API-Gateway. Moreover, changes to the API contract can be easily managed by the API-Gateway, which is responsible for serving the content to different consumers, providing only the data they need. |
| **Lust** (Bryant [13]) **Focus on Technology** (Richardson [15]) | Usage of the latest technologies |
| **Gluttony** (Bryant [13]) | Usage of too many different communication protocols such as HTTP, ProtoBuffs, Thrift, etc. |
| **Greed** (Bryant [13]) | All the services belong to the same team. |
| **Sloth** (Bryant [13]) | Creation of a distributed monolith due to the lack of independence of microservices |
| **Wrath** (Bryant [13]) **Magic Pixiedust** (Richardson [15]) | Believing a sprinkle of microservices will solve the development problems |
| **Microservices as the Goal** (Richardson [15]) | Migrating to microservices because everybody do it, and not because the company need it |
| **Scattershot Adoption** (Richardson [15]) | Multiple teams independently adopting microservices without coordination. |
| **Envy** (Bryant [13]) **The More the Merrier** (Richardson [15]) | Creating as many microservices as possible. |
| **Trying to fly before you can walk** (Richardson [15]) | Migrating to microservices while lacking the key skills, e.g. clean code object-oriented design automated testing |
| **Pride** (Bryant [13]) | Testing in the world of transience |
| **Red Flag Law** (Richardson [15]) | Adopting microservices without changing process, policies and organization |

Unlike these works, we identified a set of microservices anti-patterns based on bad practices reported by the 72 participants of our previous survey [6] and on the 27 participants of this current study. In the Results Section, we map our set of microservices anti-pattern to the bad practices identified in Table 1.7.

## 1.5 Conclusion

In this work, we identified a set of 20 microservices anti-patterns based on bad practices experienced by practitioners while developing microservices-based systems. This review, based partly on earlier work, has resulted as a consequence of the additional surveys in a significantly more comprehensive and up-to-date catalogue of patterns. Furthermore, we identified change in perception over the years that the microservice architectural style is in use.

The results show that splitting a monolith, including splitting the connected data and libraries, is the most critical issue, resulting in potential maintenance issues when the cuts are not done properly. Moreover, the complexity of a distributed system increases the system complexity, especially when dealing with connected services that need to be highly decoupled from any point of view, including communication and architecture (Hardcoded Endpoints, No API-Gateway, Inappropriate Service Intimacy, Cyclic Dependency).

This work resulted in the following five lessons learned:

- **Lesson learned 1**: Besides traditional anti-patterns, microservices-specific anti-patterns can also be problematic for the development and maintenance of microservices-based systems. Developers can already benefit from our catalog by learning how to avoid experiencing the related bad practices bot from an organizational and an architectural point of view.
- **Lesson learned 2**: Splitting a monolith into microservices is about identifying independent business processes that can be isolated from the monolith and not only about extracting features in different web services.
- **Lesson learned 3**: The connections between microservices, including the connections to private data and shared libraries, must be carefully analyzed.
- **Lesson learned 4**: As a general rule, developers should be alerted if they need to have deep knowledge of the internal details of other services or if changes in a microservice require changes in another microservice.

The proposed taxonomy of anti-patterns can be used by practitioners as a guideline to avoid the same problems happening to them as faced by our interviewees. Moreover, the catalog is also a starting point for additional research on microservices. It is important to note that, even though the identified anti-patterns reflect the opinion of the interviewed developers, the rating of the harmfulness of the reported anti-patterns is only based on the perception of the practitioners and needs to be empirically validated.

Microservice is still a very recent technology and future long-term investigation will be needed to evaluate the harmfulness and the comprehensiveness of our catalog. This, together with more in-depth empirical studies (such as controlled experiments), will be part of our future work.